\documentclass{article}

\usepackage{arxiv}
\renewcommand{\shorttitle}{Toward Improved Operational Efficiency and Sustainability}
\usepackage{xcolor}
\usepackage{graphicx}
\usepackage{subcaption}
\usepackage{hyperref}

\usepackage[utf8]{inputenc} % allow utf-8 input
\usepackage[T1]{fontenc}    % use 8-bit T1 fonts
\usepackage{hyperref}       % hyperlinks
\usepackage{url}            % simple URL typesetting
\usepackage{booktabs}       % professional-quality tables
\usepackage{amsfonts}       % blackboard math symbols
\usepackage{nicefrac}       % compact symbols for 1/2, etc.
\usepackage{microtype}      % microtypography
\usepackage{lipsum}		% Can be removed after putting your text content
\usepackage{graphicx}
\usepackage[numbers]{natbib}
\usepackage{doi}

\hypersetup{
pdftitle={A template for the arxiv style},
pdfsubject={q-bio.NC, q-bio.QM},
pdfauthor={David S.~Hippocampus, Elias D.~Striatum},
pdfkeywords={First keyword, Second keyword, More},
}

\title{Initial Field Test of a Quadrupedal Robot at SLAC Beam Dump East~(BDE): A Sandbox for Robotics in Accelerator Operations}
\author{M. Bai, M. A. Montironi, A. Saad, T. Thayer\thanks{This work is supported by the SLAC resources including computational resources}\\
SLAC National Accelerator Laboratory}
\begin{document}

\maketitle

\begin{abstract}
Deploying field robotics in large-scale particle accelerator facilities offers significant potential to enhance operational efficiency, promote facility sustainability, and reduce personnel exposure to radiation and other hazards. As key underlying technologies mature, field robotics are seeing broader deployment across various industrial sectors. However, accelerator facilities have remained cautious in adopting these systems, primarily due to concerns regarding cost versus return on investment and the risk of equipment collisions. In addition, operational experience with SLAC's Remotely Operated Accelerator Monitoring platform, ROAM~\cite{roam} highlighted critical technical challenges, including radiation tolerance and poor network connectivity in enclosed accelerator areas ~\cite{WE3AO06}. This paper presents the findings from a recent test drive of a Boston Dynamics SPOT quadrupedal robot—equipped with a SPOT Arm payload—conducted at SLAC's Beam Dump East (BDE) facility. We evaluate the technical feasibility and operational considerations of deploying quadrupedal field robotics within accelerator environments. Furthermore, we report key practical challenges, such as protective suits that impede navigation sensors and low-light infrared navigation in unlit spaces. Finally, we outline potential development pathways toward adaptive learning and autonomous execution for future routine field robotics operations in accelerator facilities.

\end{abstract}

\section{Introduction and Motivation}
Routine search, inspection, and maintenance procedures in high-energy particle accelerator facilities traditionally require personnel to enter enclosed tunnels, beamlines, and radiologically controlled areas. While these practices have proven effective over the past several decades, the required operational effort grows substantially as accelerator facilities expand in both scale and technical complexity. Recent decadal tracking of the High Energy Physics budget distribution across research, facilities, and projects indicates the increase of facility operation~\cite{Rameika:2026fcc}. This continuous expansion drives up operational costs significantly. With constrained budgets, developing methods to maintain operational expenditure at sustainable levels is essential to preserving funding for core research and development, which ultimately accelerates scientific discovery and technological advancement.

In addition to cost sustainability, traditional maintenance models necessitate periodic facility shutdowns, both for scheduled maintenance and in response to unexpected technical issues. In the case of unscheduled interventions, the typical workflow requires stopping accelerator operations, waiting for radiation levels to cool down to acceptable thresholds, and deploying subject matter experts (SMEs) to inspect and repair the equipment/system. In many instances, once the machine is powered down, initial investigations reveal no observable errors. Consequently, the ability to conduct remote monitoring and diagnostics while the machine remains operational could substantially enhance operational efficiency, maximizing beam time available for science users. Considering that high-energy particle accelerators operate 24/7, the effects of such unplanned interventions are particularly detrimental during off-hours operations when SMEs are generally not immediately available. Furthermore, because accelerator environments present elevated radiation levels alongside physical hazards, such as high voltage equipment and confined, highly congested legacy spaces, current practices inevitably expose personnel to safety risks. Compliance with the resulting multi-layered regulatory safeguards places additional demands on facility resources and operational throughput.

Transitioning these tasks to autonomous and/or teleoperated field robotics could directly address two critical imperatives:
\begin{itemize}
    \item Operational Efficiency and Sustainability: Continuous, objective, and reconfigurable data acquisition during robotic missions complement human observations with higher-frequency, consistently acquired data on machine and environmental conditions, enabling condition-based and predictive maintenance while reducing facility downtime.
    \item Environmental, Radiation Safety, and Hazard Mitigation: Autonomous or teleoperated field robotic inspection reduces the need for human presence in hazardous environments and areas with elevated radiation levels.
\end{itemize}

\subsection*{In-House Robotic Platforms for Autonomous and Remote Monitoring at SLAC}
Motivated by these imperatives, motion control experts at SLAC initiated the development of custom wheeled robotic platforms constructed from commercial off-the-shelf components and open-source software frameworks. Two Remotely Operated Accelerator Monitoring (ROAM) custom mobile platforms equipped with specialized sensor suites were constructed \cite{roam} \cite{WE3AO06} to perform radiation mapping, environmental surveillance, and visual inspection within accelerator housing structures. These systems were evaluated across multiple locations in the SLAC linear accelerator complex, yielding the following operational observations and accomplishments:
\begin{itemize}
    \item ROAM established baseline requirements for remote mobility, tethered/wireless communication, and radiation-tolerant sensor payloads.
    \item Radiation Susceptibility: Onboard commercial electronics exhibited vulnerability to radiation-induced degradation during prolonged tunnel exposure.
    \item Network Bandwidth Constraints: Wireless signal attenuation and bandwidth limitations in subterranean accelerator housing constrained real-time high-definition video streaming.
\end{itemize}
Despite these initial encouraging results, routine deployment of the ROAM platforms was not sustained—a trend mirrored across other accelerator facilities despite rapid advancements in commercial autonomous navigation. Widespread adoption of field robotics in accelerator environments has primarily been hindered by two main concerns:
\begin{itemize}
    \item Cost vs. Return on Investment: While initial integration costs for specialized robotics can be significant, return on investment is maximized when core navigation, inspection, and data-processing frameworks are shared across multiple beamlines and user facilities.
    \item Risk of Equipment Collision: Accelerator tunnels host sensitive, high-value apparatus (e.g., cryomodules, RF cavities, precision magnets) demanding strict zero-collision guarantees. Validating robotic navigation in representative, decommissioned environments—such as SLAC’s Beam Dump East facility—provides essential real-world data to evaluate obstacle detection and avoidance capabilities prior to active tunnel deployment.
\end{itemize}
\section{Recent Test Voyage of SPOT at Beam Dump East (BDE)}
In 2023, SLAC Facility Operations procured a Boston Dynamics SPOT quadrupedal robot equipped with an articulated 5th-arm payload, with the initial objective of enabling SPOT to open high-voltage breakers as part of standard certification procedures. Recently, SPOT was made available to the Accelerator Directorate. Concurrently, End Station A (ESA) has resumed scientific operations following the successful re-establishment of a multi-GeV electron beam delivered from the SLAC superconducting LINAC to ESA~\cite{LESA}. Consequently, the entire ESA area must adhere to accelerator operational safety procedures, which mandate search sweeps during each intervention in beam operations to prevent personnel radiation exposure.

\begin{figure}
        \centering
        \includegraphics[width=0.5\linewidth]{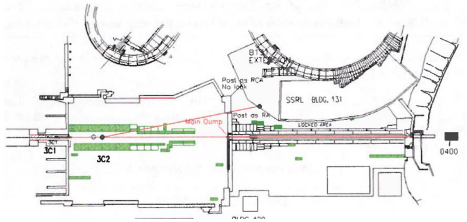}
        \includegraphics[width=0.5\linewidth]{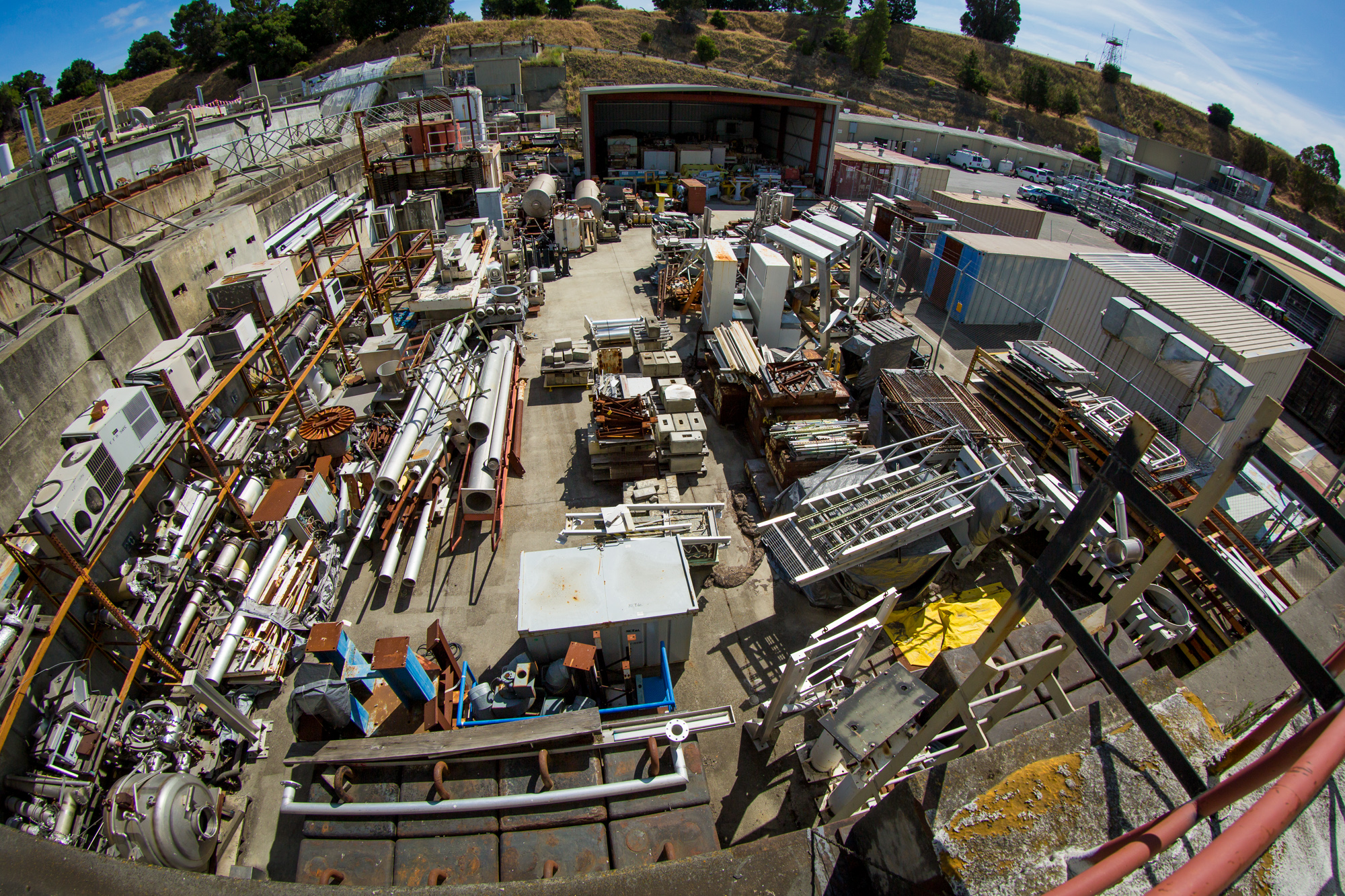}
        \caption{ESA and BDE Facility Overview. The top part of the figure is the overall layout of ESA~(on the left part of the drawing) while the lower part of the figure is the bird view of the fenced area outside the BDE tunnel corresponding to the area in the lower right part of the top part of Figure}
        \label{ESA-BDE}
    \end{figure}

As part of its legacy infrastructure, the ESA complex includes Beam Dump East (BDE), a facility historically designed to transport high-energy electron beams near the megawatt scale to a megawatt-class beam dump. Figure~\ref{ESA-BDE} illustrates the position of BDE relative to ESA. Although the high-power beam dump has been out of service for decades, the BDE line retains its classification as a Level-3 Personnel Protection Safety zone. During the past decade, the structural and electrical infrastructure has deteriorated significantly; notably, permanent lighting is absent, and the sump~(a below-grade water collection pit) area presents confined spatial access, creating physical hazards for human entry. The sump area is also classified as a radiological contamination area due to tritiated water leaking from the legacy dump. Nevertheless, as a decommissioned high-power electron beam line, BDE offers a unique opportunity to validate autonomous robotic navigation in realistic tunnel environments without risking damage to operational high-value accelerator equipment. This environment provides valuable empirical data for evaluating obstacle-detection and collision-avoidance capabilities prior to active tunnel deployment.

Motivated by this opportunity, we conducted an initial test voyage with SPOT throughout BDE to evaluate quadrupedal field robotics under realistic accelerator conditions, encompassing both open floor areas and enclosed tunnel segments. Figure~\ref{ESA-BDE} illustrates the position of BDE relative to ESA.

\begin{figure}
    \centering
    \includegraphics[width=0.3\linewidth]{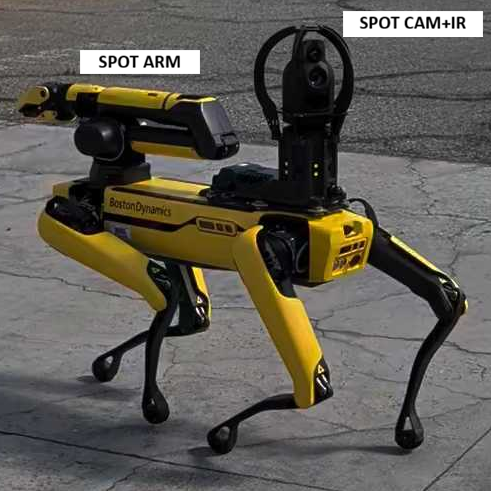}
    \caption{SPOT robot configuration used for the test. The SPOT Arm and Cam+IR payloads are labeled.}
    \label{fig:SPOT-with-payload}
\end{figure}

The SPOT robot deployed for this application was equipped with a "SPOT Arm" for manipulation and a "SPOT Cam+IR" payload composed of visible-light and Infrared (IR) cameras mounted on a Pan-Zoom-Tilt (PZT) stage (Figure ~\ref{fig:SPOT-with-payload}). Both the arm and the camera payload were mounted on the back of the robot and were used during the test to collect sample images and to attempt opening doors. The SPOT robot was controlled by an operator watching the live feed from the robot's cameras displayed on the SPOT APP UI on a hand-held controller. A local wireless network connection between the robot and the hand-held controller allowed the mission to proceed even without site WiFi availability, but constrained the operator to remain close to the robot. Locomotion and obstacle avoidance were managed by the robot's proprietary software. Although the available payload and software are capable to deploy the robot on "autowalk" missions, this feature was not tested during the initial deployment. The robot manufacturer defines "autowalk" as the ability to record and replay autonomous missions along a pre-planned path with pre-defined data acquisition locations.  

\begin{figure}
    \centering
    \includegraphics[width=0.3\linewidth]{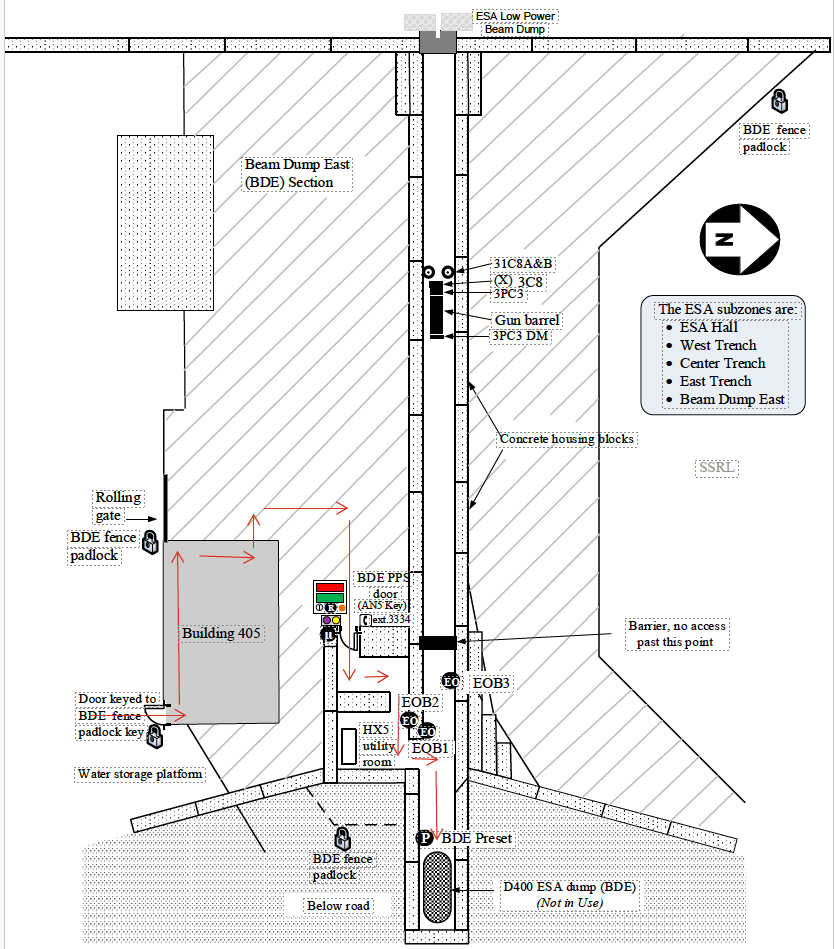}
    \caption{BDE floor plan. The red arrows correspond to the route traveled by SPOT.}
    \label{fig:SPOT-BDE-Route}
\end{figure}

The test voyage spanned approximately 33 minutes, during which SPOT was operated via its primary control interface. Figure~\ref{fig:SPOT-BDE-Route} shows the route of SPOT's maiden voyage.
Key highlights and operational accomplishments include:
\begin{itemize}
    \item Full Area Traversal: SPOT successfully traversed the complete BDE facility and narrow underground tunnel without collisions, requiring minimal manual clearance assistance from team members.
    \item Uneven Terrain Navigation: SPOT demonstrated robust locomotion stability across uneven flooring, scattered debris, and floor height transitions as shown in Figure~\ref{fig:uneventerrain} 
    \begin{figure}[htbp]
        \centering
        \begin{subfigure}{0.48\linewidth}
            \centering
                \includegraphics[width=0.5\linewidth]{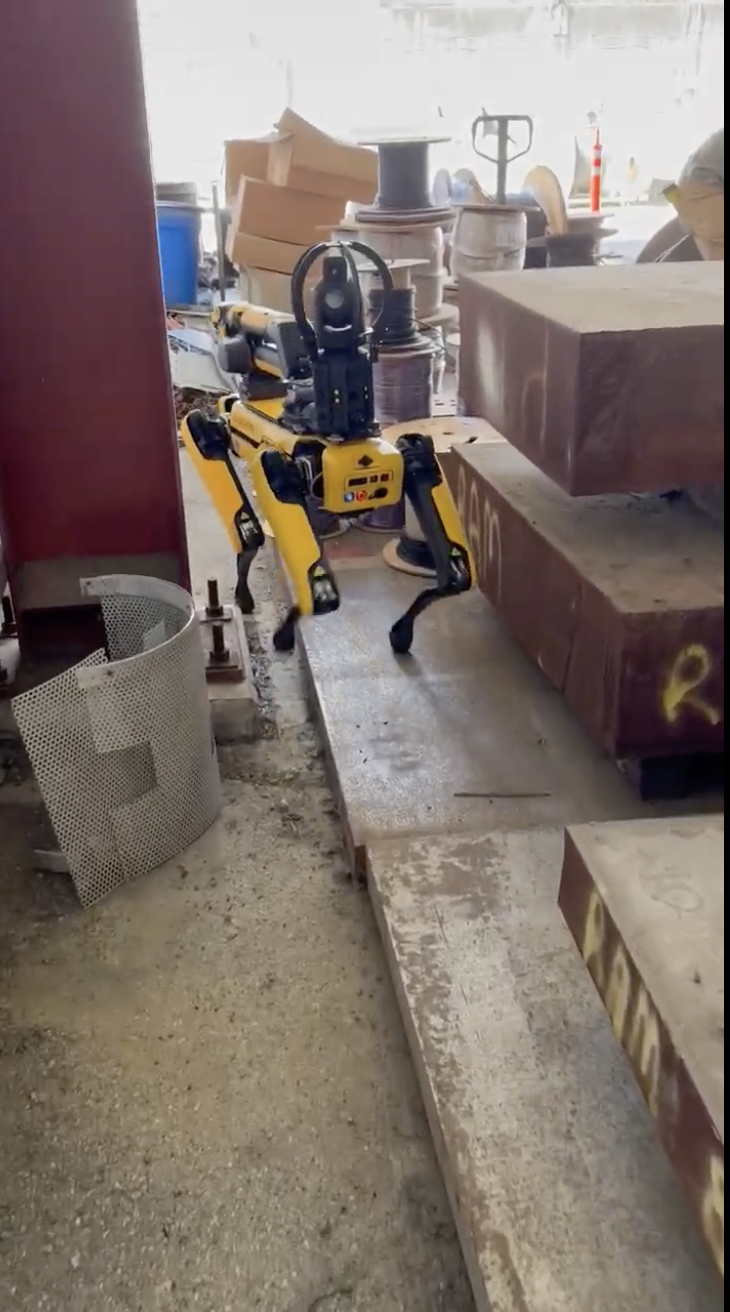}
                \caption{SPOT navigating through Building 405, where uneven surfaces and stored equipment constrain the available path.}
        \end{subfigure} \hfill 
        \begin{subfigure}{0.48\textwidth} 
            \centering 
                \includegraphics[width=0.55\linewidth]{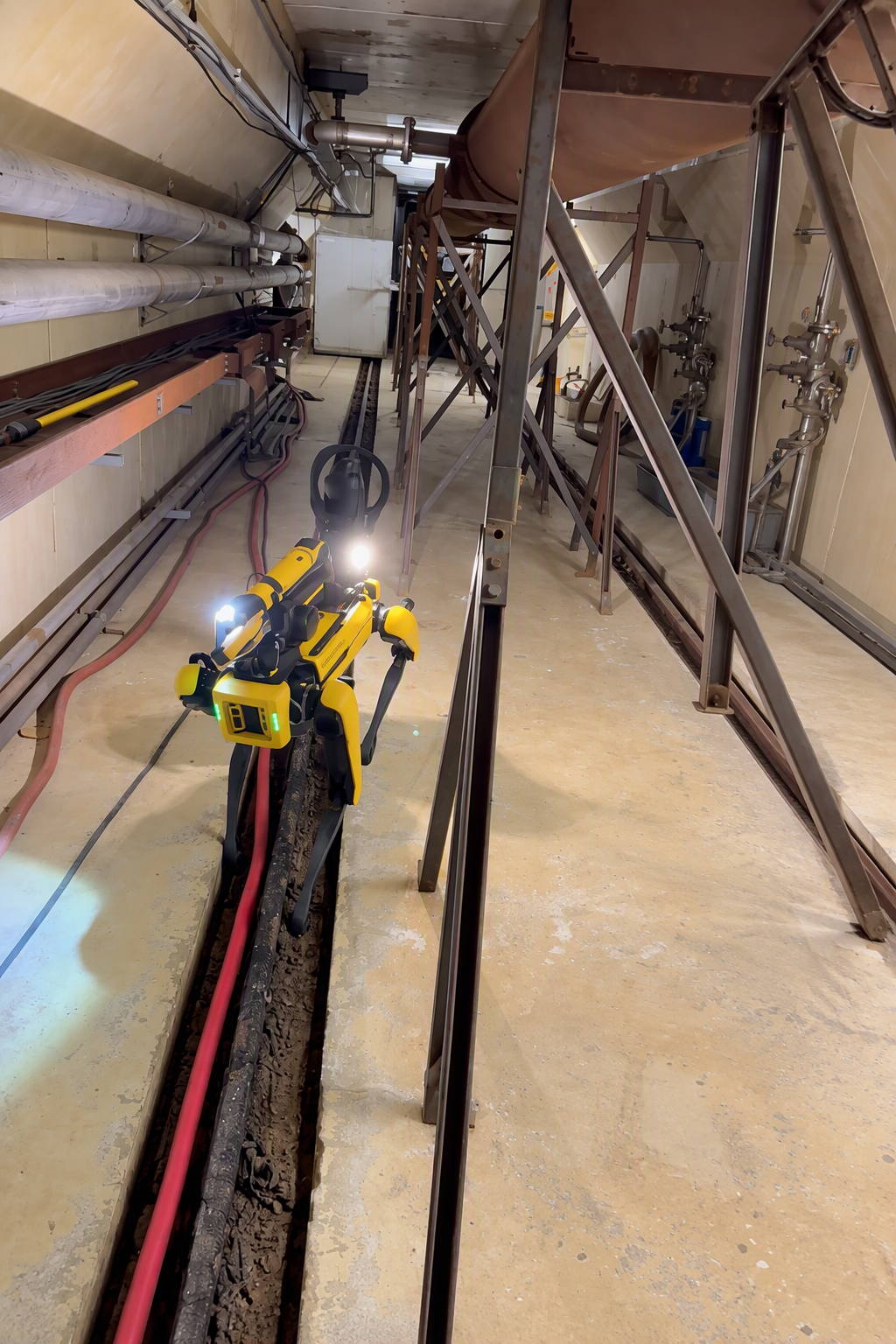} 
                \caption{SPOT navigating along the BDE tunnel, where uneven surfaces and multiple obstructions are present.}
        \end{subfigure}
        \caption{SPOT navigating uneven and obstructed surfaces in Bldg 405 and along the BDE tunnel toward the sump area.}
        \label{fig:uneventerrain}
    \end{figure}
    \item Targeted Inspection Capabilities: SPOT maneuvered adjacent to the platform facing the beam dump and deployed its 5th-arm payload to capture high-resolution visual and thermal inspection data of the sump area as shown in Figure~\ref{fig:atsump}
     \begin{figure}[htbp]
        \centering
        \begin{subfigure}{0.48\textwidth} 
        \centering
                \includegraphics[width=0.5\linewidth]{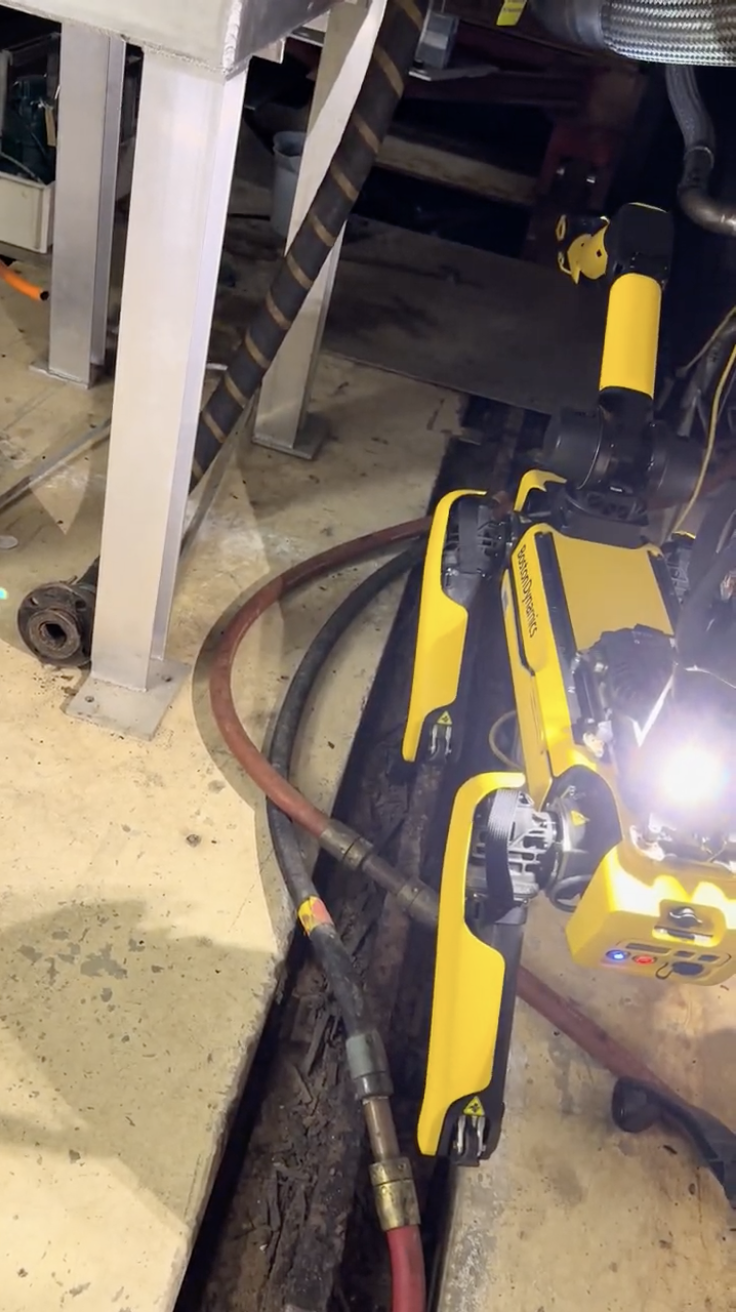}
                %\caption{positioned in front of SUMP and exercise the 5th load to explore the possibility of mapping the area with either camera and other sensors to determine any presence of person}
        \end{subfigure} \hfill 
        \begin{subfigure}{0.48\textwidth} 
            \centering
            \includegraphics[width=0.5\linewidth]{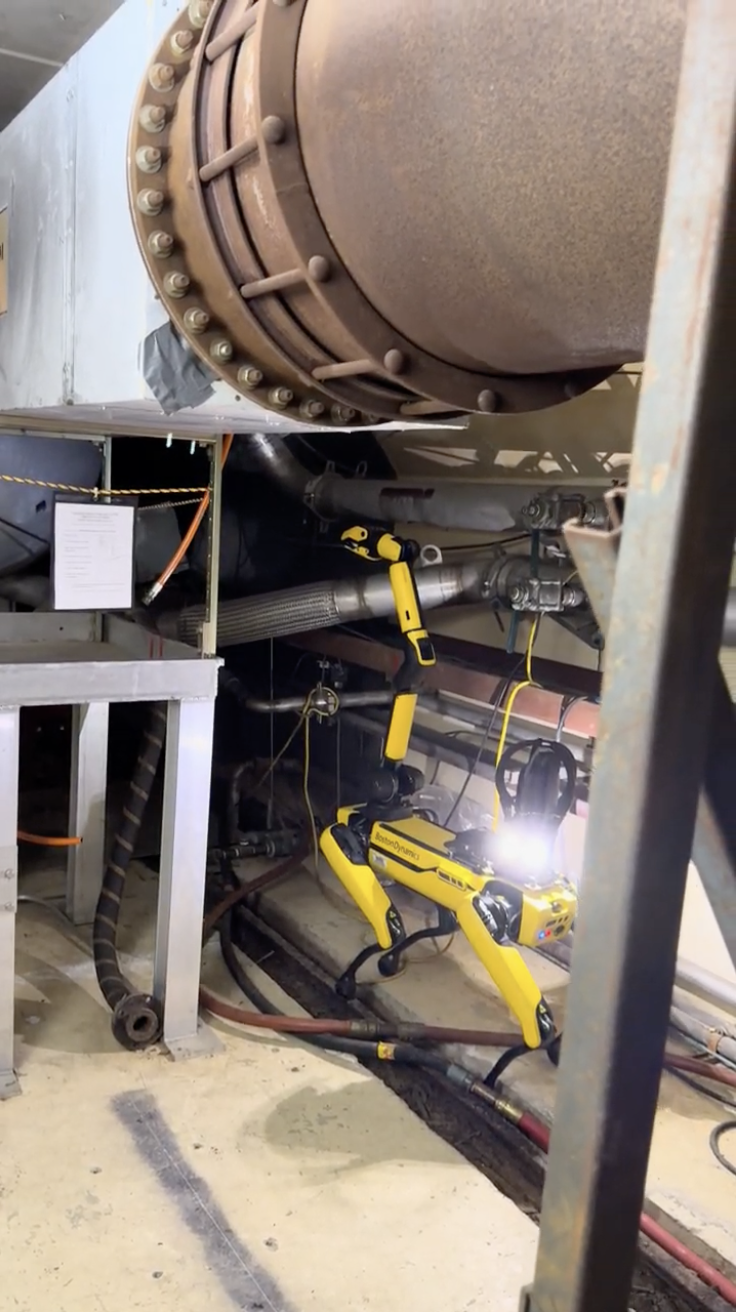}
            %\caption{positioned in front of SUMP and exercise the 5th load to explore the possibility of mapping the area with either camera and other sensors to determine any presence of person}
             \end{subfigure}
        \caption{SPOT positioned in front of the BDE sump area, exercising the arm payload, which can be equipped with a camera or other sensors to explore the potential for mapping the area and detecting the presence of personnel.}
        \label{fig:atsump}
    \end{figure}
\end{itemize}
While this initial BDE test drive confirmed baseline feasibility, several key technical challenges were identified for ongoing R\&D prior to full operational adoption:
\begin{itemize}
    \item Physical and Sensor Integration
    \begin{itemize}
        \item Protective Suit Redesign: Protective suits deployed to shield the robot in contaminated or dusty environments inadvertently obstructed onboard optical and navigation sensors. Future suit revisions must integrate multi-spectral sensor windows to maintain full navigation functionality.
        \item Spatial Envelope and Clearance Margins: SPOT's default obstacle-avoidance software enforces conservative spatial clearance envelopes. Targeted guidance was needed to avoid an area with meshed metal flooring. An ad hoc ramp was also assembled on the spot to enable SPOT to safely cross an oversized barrier. In short, further development and validation of the robotic navigation capabilities are needed for real-world deployment.
        \item Low-Light/IR Navigation in Confined non-Illuminated Spaces: Despite integrated illumination, SPOT encountered degraded visual perception in narrow, unlit passages. This affected the ability of the operator to remotely drive the robot through the control interface. Additional lights had to be introduced to successfully navigate the environment. Upgraded multi-spectral active lighting and/or supplementary LiDAR integration are recommended.
    \end{itemize}
    \item Control and Data Architecture
    \begin{itemize}
        \item Controller Independence and Autonomy: The current deployment relied on real-time teleoperation through a hand-held controller. Full facility integration requires autonomous mission execution relying on GPS-denied mapping and adaptive path planning.
        \item Remote Deployment and Data Management: Sensor data collected during missions was stored locally on the hand-held controller. A centralized software application is needed to remotely trigger missions, stream the data in real time, and deposit inspection data into facility databases for automated anomaly detection and long-term trend analysis.
        \item Network Connectivity: The lack of electrical power in the BDE area also resulted in a lack of network (WiFi) connectivity. Full facility integration requires a robotic platform able to continue operating autonomously while offline in between locations with known network connectivity.
    \end{itemize}
\end{itemize}
\section{Conclusion and Future R\&D Roadmap}
The initial test voyage at SLAC's BDE facility demonstrates the technical feasibility and potential of deploying quadrupedal field robotics for routine accelerator searches and environmental monitoring. The trial demonstrated that modern legged platforms, such as SPOT, possess the requisite mobility and agility to navigate complex, cluttered accelerator environments, including narrow passages and uneven terrain, without collisions during this initial test.

To mitigate cost and return-on-investment concerns, we propose two complementary strategies. First, cross-institutional collaboration across U.S. Department of Energy (DOE) accelerator facilities can facilitate the sharing of core software frameworks and operational workflows. Second, integrating adaptive online learning enables field robots to autonomously adapt to specific tunnel environments, reducing reliance on manual pre-planning.

In addition, technical limitations related to radiation degradation and connectivity constraints must be systematically addressed.
 
Accordingly, future R\&D efforts will focus on:
\begin{itemize}
    \item Developing an adaptive learning framework for field robotics in accelerator environments.
    \item Establishing a unified, remote mission control architecture.
    \item Quantitatively assessing the radiation tolerance of key robotic sub-systems.
    \item Integrating radiation-hardened sensor payloads to extend operational longevity.
    \item Exploring subterranean mobile network enhancements for reliable telemetry.
\end{itemize}
\section{Acknowledgment}

\end{document}